\documentclass[amsmath,amssymb,aps]{revtex4-2}

\usepackage{xcolor,graphicx}        
\usepackage{mwe}
\usepackage{float}
\usepackage{subfig}
\usepackage{amsmath,amsfonts,amssymb}
\usepackage{tikz}
\usepackage{braket}
\usepackage{hyperref}
\hypersetup{
	colorlinks = true,
	linkcolor = red,
	anchorcolor = blue,
	citecolor = blue,
	filecolor = blue,
	urlcolor = blue}
\usepackage[top=2cm,bottom=2cm,right=2cm,left=2cm]{geometry}

\begin{document}

\title{Complete Rabi oscillations in the ion–laser interaction}

\author{M. A. García-Márquez$^{1*}$, H. M. Moya-Cessa, I. Ramos-Prieto, F. Soto-Eguibar }
\affiliation{Instituto Nacional de Astrof\'isica, \'Optica y Electr\'onica, Tonantzintla, Puebla.}





\begin{abstract}
We present the dynamics of a single harmonically trapped ion interacting with a laser, considering a linear combination of two eigenstates of the system as the initial state. The conditions on the physical parameters that allow for the evolution of the system are discussed. We are able to obtain analytical results even though no approximations are realized and are able to show Rabi oscillations for such initial states.
\end{abstract}

\maketitle




\section{Introduction}
The study of the interaction between a trapped ion and a laser beam \cite{Moya-Cessa et al. (2012)} has been of great importance due to the ease with which trapped ions may be manipulated and controlled, enabling the generation of nonclassical states of the trapped ion's vibrational motion \cite{Meekhof et al. (1996), Blatt et al. (1995), Moya-Cessa y Tombesi (2000), de Matos Filho y Vogel (1996a), de Matos Filho y Vogel (1996b), Monroe et al. (1996),Ramos-Prieto et al. (2020)}. Furthermore, this control capability has led to trapped ions being considered one of the foremost platforms for implementing quantum simulators \cite{Blatt y Roos (2012)}, which allow, for example, proposals to study the quantum Rabi model across all parameter regimes \cite{Pedernales et al. (2015)} or to produce a fast quantum Rabi model \cite{H. M. Moya-Cessa (2016)}. In addition, trapped ions have been proposed for various applications, such as quantum computing \cite{Cirac y Zoller (1995)} and precision spectroscopy \cite{Wineland et al. (1992)}.

Due to the non-linearity of the Hamiltonian describing the interaction between the trapped ion and the laser, various approaches exist to solve the system's dynamics. For instance, the Lamb-Dicke approximation assumes that the ion is confined to a region much smaller than the laser wavelength. Other methods involve unitary transformations that linearize the Hamiltonian \cite{H. M. Moya-Cessa et al. (1999)},  resulting in a Jaynes-Cummings-type Hamiltonian that may be solved analytically using the rotating wave approximation (RWA) under the condition of the intensity regime $\Omega\approx \nu$ where $\Omega$ is the field intensity and $\nu$ is the ion's vibrational frequency. Subsequently, transformations on the linearized Hamiltonian have been employed to solve the system in different regimes, such as the use of small rotations in the resonance case ($\delta = 0$) \cite{Aguilar y Moya-Cessa (2002)}, or the use of recursive transformations without additional conditions on $\Omega$ or the Lamb-Dicke parameter $\eta$ \cite{H. M. Moya-Cessa et al. (2000)}. Moreover, Zúñiga et al. \cite{Zúñiga‐Segundo et al. (2011)} have demonstrated that all intensity regimes of the trapped ion can be analytically treated.

The eigenstates for this model have been expressed as series without a known closed-form expression \cite{H. M. Moya-Cessa et al. (2000)}. Furthermore, expressions for the eigenvalues and eigenfunctions of the Jaynes-Cummings model have been found for all coupling strengths and detunings\cite{Chen et al. (2011)}, which is relevant because the trapped ion system is equivalent to an atom interacting with a single-mode quantized electromagnetic field. 
On the other hand, a family of exact eigenstates for the ion-laser Hamiltonian, also known as trapping states \cite{Moya-Cessa et al. (2003), Ramos Prieto y Moya-Cessa (2017)}, which are valid only under an appropriate combination of the system's physical parameters, has been identified. However, since these states do not constitute a complete basis, an exact solution can only be found for the eigenstates belonging to this family.

In this work, the dynamics of a single harmonically trapped ion interacting with a laser is studied, considering as the initial state the sum of two eigenstates found in \cite{Moya-Cessa et al. (2003)}, with particular attention given to the conditions of the physical parameters that allow the system's evolution. Additionally, the atomic inversion of the system and the average phonon number of the ion's vibrational motion are calculated.

\section{Ion–laser interaction in a trap}
The Hamiltonian describing a trapped ion interacting with a laser is given by
\begin{equation}
\hat{H}=\hat{H}_{\text{vib}}+\hat{H}_\text{at}+\hat{H}_\text{int},
\label{Hamiltonian of the trapped ion}
\end{equation}
where $\hat{H}_\text{vib}$ represents the vibrational energy of the ion's center of mass, whose vibrational motion can be approximated as that of a harmonic oscillator, $\hat{H}_\text{at}$ denotes the internal energy of the ion, which will be modeled as a two-level system, and $\hat{H}_\text{int}$ describes the interaction energy between the ion and the laser \cite{Moya-Cessa et al. (2012)}. The dipole approximation is used in the interaction between the ion and the laser, that is, the interaction will be given by $-e\vec{r}\cdot\vec{E}$, where $-e\vec{r}$ is the dipole moment of the ion and $\vec{E}$ is the electric field of the laser, which will be considered as a plane wave. Using the optical rotating wave approximation, the Hamiltonian is then written as (we set for simplicity $\hbar=1$)
\begin{equation}
    \hat{H}=\nu\hat{n}+\frac{\omega_{21}}{2}\hat{\sigma}_z+\Omega\left[ e^{i(k\hat{x}-\omega t )} \hat{\sigma}_{+}+ e^{-i(k\hat{x}-\omega t )} \hat{\sigma}_{-}\right],
    \label{Hamiltonian of the trapped ion 2}
\end{equation}
where $\nu$ is the frequency of the vibrational motion, $\omega_{21}$ is the transition frequency between the ground state and the excited state of the ion, $\Omega$ is the Rabi frequency, $\omega$ and $k$ are the frequency and wave number of the electric field of the laser,   $\hat{n}$ is the number operator, and $\hat{\sigma}_{z}$, $\hat{\sigma}_{+}$ y $\hat{\sigma}_{-}$ are the Pauli matrices. The vibrational Hamiltonian has been shifted by the vacuum energy.

The Hamiltonian \eqref{Hamiltonian of the trapped ion 2} can be written as \cite{Zúñiga‐Segundo et al. (2011)}
\begin{equation}
    \hat{H}_\text{ion}=\nu\hat{n}+\frac{\delta}{2}\hat{\sigma}_z+\Omega \left[  \hat{\sigma}_{+}\hat{D}(i\eta)+\hat{\sigma}_{-}\hat{D}^{\dagger}(i\eta)\right],
    \label{Hamiltonian of the trapped ion 3}
\end{equation}
where $\eta = k\sqrt{\frac{1}{2\nu }} $ is the (real) Lamb-Dicke parameter, $ \delta = \omega_{21}-\omega$ is the laser-ion detuning and 
\begin{equation}
    \hat{D}(i\eta)= e^{i\eta (\hat{a}+\hat{a}^{\dagger})} 
    \label{Displacement operator}
\end{equation}
is the Glauber  displacement operator \cite{Glauber}.

\subsection{Eigenstates of the system}
\begin{figure}[h!]
    \centering
    \includegraphics[width=.5\linewidth]{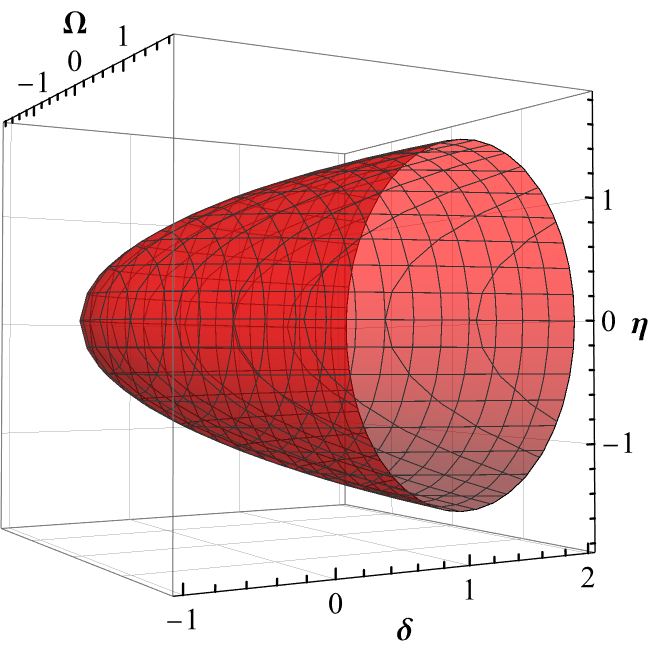}
    \caption{Surface corresponding to the  condition (\ref{Condition 0+}) with $\nu=1$, for which the eigenfunction $|\psi_0^{+}\rangle$ is valid.}
    \label{fig:Condition 0+ xyz}
\end{figure}

Under certain conditions on the physical parameters $\nu$, $\delta$, $\Omega$ and $\eta$ of the system, it is possible to find exact eigenstates for the Hamiltonian of the single trapped ion, given in Eq. (\ref{Hamiltonian of the trapped ion 3}), which can be written as a finite sum of number states and displaced number states \cite{Moya-Cessa et al. (2003)}. One of the basic eigenstates of this system, in the atomic basis and with its respective eigenvalue and condition, is
\begin{equation}
    |\psi_0^{+}\rangle=\frac{1}{N_0^{+}} 
    \begin{pmatrix}
    \frac{\Omega}{\nu}|0\rangle +\frac{\nu}{\Omega}i\eta |1\rangle \\
    |-i\eta \rangle
    \end{pmatrix}, \quad \quad
    E_0^{+}=\nu +\frac{\delta}{2},
    \label{Eigenfunction 0+}
\end{equation}
where $N_0^{+}= \left[1+\frac{\Omega^2}{\nu ^2}+\left(\frac{\eta \nu}{\Omega}\right)^2\right]^{1/2}$ is the normalization constant and $|-i\eta \rangle$ is the coherent state $|-i\eta \rangle=\hat{D}^\dagger\left(i\eta\right)|0\rangle$. This ket is a valid eigenket when the condition 
\begin{equation}
    \frac{\Omega^2}{\nu^2}+\eta^2-\frac{\delta}{\nu}=1
    \label{Condition 0+}
\end{equation}
is satisfied. 
Figure \ref{fig:Condition 0+ xyz} shows the surface that represents the condition (\ref{Condition 0+}) with $\nu=1$. Any point on the surface represents a suitable selection of the values of the physical parameters that ensure that the eigenfunction $|\psi_0^{+}\rangle$ is valid.

\begin{figure}[h!]
     \centering
     \includegraphics[width=.5\linewidth]{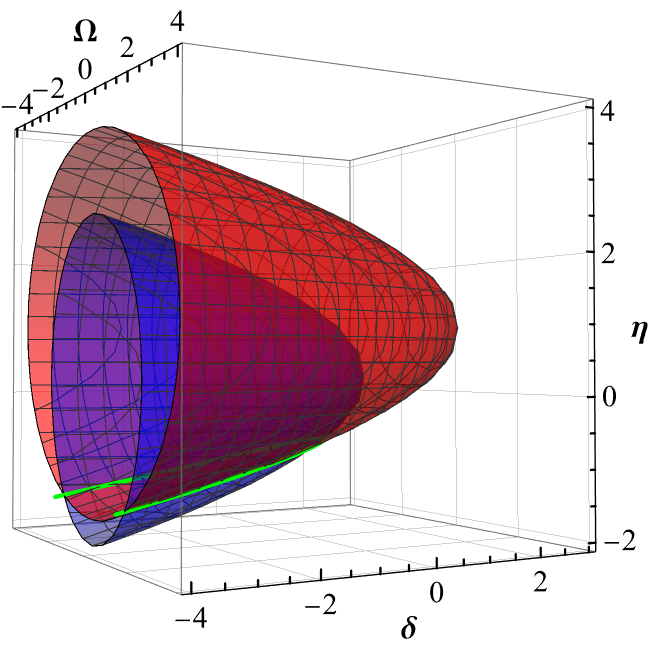}
     \caption{
     The surfaces shown give rise to the regions where the eigenket $|\psi_1^{-}\rangle$ is allowed. The first condition from equation (\ref{Conditions 1-}) is shown in red, the second condition is shown in blue and the curve where the eigenket $|\psi_1^{-}\rangle$ is valid is shown in green, with $\nu=1$ and $m=-2i$. This curve is parameterized by equations (\ref{Conditions 1- z}) and (\ref{Conditions 1- y}).}
     \label{fig:Conditions 1+ xyz}
\end{figure}

Another eigenket $|\psi_1^{-}\rangle$, with its corresponding eigenvalue $E_1^{-}$, is 
\begin{eqnarray}
     |\psi_1^{-}\rangle=\frac{1}{N_1^{-}}
     \begin{pmatrix}
d_0 | i\eta \rangle + d_1 | i\eta ;1 \rangle \\
\frac{\Omega}{\nu} \left( \frac{d_0}{2} | 0 \rangle + d_1 | 1 \rangle - i \eta \frac{\nu^2}{\Omega^2}\sqrt{2} d_1 | 2 \rangle \right)
\end{pmatrix}, \qquad E_1^{-} = 2\nu - \frac{\delta}{2}, 
\label{Eigenfunction 1-}
\end{eqnarray}
where $d_0$ and $d_1$ are complex constants, $N_1^{-}=\left\{\left[1+\left(\frac{\Omega}{2\nu}\right)^2\right] |d_0|^2 +\left[1+\left(\frac{\Omega}{\nu}\right)^2+2\left(\eta \frac{\nu}{\Omega}\right)^2\right]|d_1|^2\right\}^{\frac{1}{2}} $ is the normalization constant and $| i\eta ;1 \rangle=\hat{D}(i\eta)|1\rangle$ is a displaced number state.
This eigenket is valid when it satisfies the following conditions:
\begin{equation}
2-\eta^2-\frac{\delta}{\nu}-\frac{\Omega^2}{2\nu^2}=-i \eta m , \quad \quad \quad \quad  1-\eta^2-\frac{\delta}{\nu}-\frac{\Omega^2}{\nu^2}=\frac{i\eta}{m}, 
    \label{Conditions 1-}
\end{equation}
where $m=\frac{d_1}{d_0}$. From Eq. \eqref{Conditions 1-}, it follows that $m$ must be a purely imaginary number since $\nu$, $\Omega$, $\delta$, $\eta$ are real numbers. Combining the equations in (\ref{Conditions 1-}), it is easily observed that
\begin{equation}
\eta=\frac{im\left(1+\frac{\Omega^2}{2\nu^2}\right)}{m^2+1}, 
\label{Conditions 1- z}
\end{equation}
and 
\begin{eqnarray}
        \delta=\nu\left(1-\frac{\Omega^2}{\nu^2}-\eta^2-i\frac{\eta}{m}\right)= \nu \left(1-\frac{\Omega^2}{\nu^2}+\frac{m^2\left(1+\frac{\Omega^2}{2\nu^2}\right)^2}{(m^2+1)^2}+\frac{1+\frac{\Omega^2}{2\nu^2}}{m^2+1}\right);
    \label{Conditions 1- y}
\end{eqnarray}
that is, given the values of $\nu$ and $m$, there will be a parametric curve, with $\Omega$ as the parameter, where both conditions in Eq. (\ref{Conditions 1-}) are satisfied simultaneously. Figure \ref{fig:Conditions 1+ xyz} shows the curve parameterized by equations (\ref{Conditions 1- z}) and (\ref{Conditions 1- y}) and the two surfaces given by the conditions in equation (\ref{Conditions 1-}), with $\nu=1$ and $m=-2i$. Any point in the curve represents a suitable selection of the values of the physical parameters that ensure that the ket $|\psi_1^{-}\rangle$ is a valid eigenket. 

\section{Conditions for the evolution of the initial state 
$|\psi(0)\rangle=\frac{1}{N}\left[c_1|\psi_0^{+}\rangle +c_2 |\psi_1^{-}\rangle \right] $}
The initial condition is proposed as a combination of the eigenkets $|\psi_0^{+}\rangle $ and $|\psi_1^{-}\rangle $, and is given by 
\begin{equation}
    |\psi(0)\rangle=\frac{1}{N}\left[c_1|\psi_0^{+}\rangle +c_2 |\psi_1^{-}\rangle \right],
    \label{Initial state 0+ 1-}
\end{equation}
where $c_1$, $c_2$ must satisfy that $|c_1|^2+|c_2|^2=1$ and $N$ is the normalization constant
\begin{eqnarray}\nonumber
    N = \left[ 1 + 2  \text{Re} \left\lbrace \frac{c_1^* c_2 }{N_{0}^{+*} N_{1}^-}e^{-\eta^2/2} \left( d_0 \left( \frac{3 \Omega}{2 \nu} + \frac{\nu \eta^2}{\Omega} \right) + i \eta d_1 \left( \frac{2 \Omega}{\nu} + \frac{2 \nu }{\Omega}\eta^2 - \frac{\nu}{\Omega} \right) \right) \right\rbrace \right] ^{1/2},
\end{eqnarray}
where the symbol $*$ denotes the complex conjugation operation.
The initial condition will be valid as long as the conditions on eigenkets $|\psi_0^{+}\rangle$ and $|\psi_1^{-}\rangle $, from equations (\ref{Condition 0+}) and (\ref{Conditions 1- z}),  (\ref{Conditions 1- y}) respectively, are met simultaneously.
By solving the system of equations for the parameters $\delta$, $\eta$ and $m$, the following solutions are obtained
\begin{equation}
    \eta=\pm \frac{\sqrt{3}}{2\nu}\sqrt{\nu^2-\Omega^2},\quad \delta =-\frac{\nu \eta^2}{3},\quad m=\frac{3i}{2\eta},
    \label{Solutions 0+ 1- without considering the physical properties of the parameters 1}
\end{equation}
and
\begin{equation}
    \eta=\pm\frac{1}{\nu}\sqrt{2\nu^2-\Omega^2}, \quad \delta = \nu,  \quad m= \frac{i\eta}{2}.
    \label{Solutions 0+ 1- without considering the physical properties of the parameters 2}
\end{equation}
Therefore, given $\nu$ and $\Omega$ the remaining parameters can be determined.
However, the parameters $\nu$ and $\Omega$ cannot take arbitrary values, as it must be ensured that $m$ is a purely imaginary number and that $\delta$ and $\eta$ are real numbers. We conclude that the two solutions of equation (\ref{Solutions 0+ 1- without considering the physical properties of the parameters 1}) must satisfy the following condition
\begin{equation}
    \nu^2-\Omega^2 > 0  \quad \Rightarrow   \quad \left|\Omega\right|<\nu.
    \label{Condition Omega}
\end{equation}
This leads to the allowed values of the parameters being within the following intervals
\begin{equation}
    -\frac{1}{4}<\frac{\delta}{\nu}<0,
    \label{Condition delta}
\end{equation}
\begin{equation}
    -\frac{\sqrt{3}}{2}<\eta<0 \quad \text{or} \quad 0<\eta <\frac{\sqrt{3}}{2},
    \label{Condition eta}
\end{equation}
\begin{equation}
    M<-\sqrt{3} \quad \text{or} \quad \sqrt{3}<M.
    \label{Condition M}
\end{equation}
where it has been considered that $m=\frac{d_1}{d_0}=iM$, where $M$ is a real number. Qualitatively, since $\nu>0$, for the positive solution of equation (\ref{Solutions 0+ 1- without considering the physical properties of the parameters 1})  we have that $\eta, M>0$, and for the negative solution $\eta, M<0$. It should be noted that although the intervals in which the physical parameters are constrained have been identified, it does not mean that their values can be chosen arbitrarily, as they are determined by the relations in (\ref{Solutions 0+ 1- without considering the physical properties of the parameters 1}).

We observe that the condition in equation (\ref{Condition Omega}) restricts the system to evolve in the low intensity regime, where $\nu \gg \Omega$, and in the medium intensity regime, where the vibrational frequency $\nu$ is on the order of the field intensity $\Omega$.

Now, for the two solutions of equation (\ref{Solutions 0+ 1- without considering the physical properties of the parameters 2}),  we observe that the condition $\delta = \nu$ leads to the eigenenergies corresponding to $|\psi_0^+\rangle$ and $|\psi_1^-\rangle$, being the same, i.e., there will be degenerate states with
\begin{equation}
    E_0^+= E_1^-= \frac{3}{2}\nu,
    \label{conditions on the eigenenergies -}
\end{equation}
and, therefore, there will be no dynamics in the system that can be studied.

\begin{figure}[h!]
    \centering
    \includegraphics[width=.5\linewidth]{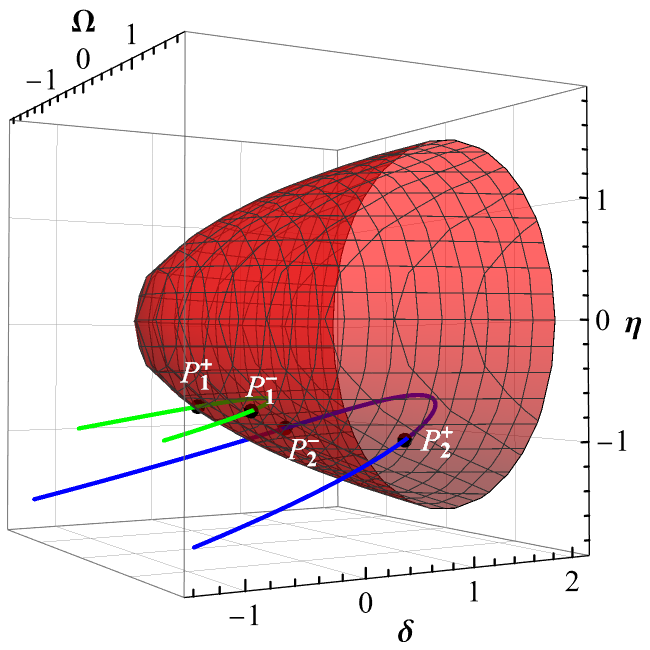}
    \caption{The surfaces shown give rise to the regions where the initial state $|\psi(0)\rangle=\frac{1}{N}\left[c_1|\psi_0^{+}\rangle +c_2 |\psi_1^{-}\rangle\right]$ is allowed. The red surface represents the condition under which eigenfunction $|\psi_0^{+}\rangle$ is valid, while the green and blue curves show the conditions under which the eigenfunction $|\psi_1^{-}\rangle $ is valid, with $m=-2 i$ and  $m=0.5 i$, respectively. Black points indicate the intersections of the conditions, and it is at these $P_1^\pm$, $P_2^\pm$ points where the initial state $|\psi(0)\rangle$ is valid for some value of $m$.}
    \label{fig:enter-label}
\end{figure}

Alternatively, by solving the system of equations in terms of $\Omega$, $\delta$, and $\eta$, it is found that
\begin{equation}
 P_{1}^\pm = (\Omega, \delta, \eta)=\left( \pm \frac{\nu}{m} \sqrt{m^2+3}, \frac{3\nu}{4 m^2}, \frac{3 i}{2 m}\right), 
\label{Solutions 0+ 1- without considering the physical properties of the parameters 1 xyz}
\end{equation}

\begin{equation}
 P_{2}^\pm =(\Omega, \delta, \eta)=\left( \pm \sqrt{2} \nu \sqrt{2 m^2+1}, \nu,2 i m\right).
\label{Solutions 0+ 1- without considering the physical properties of the parameters 2 xyz}
\end{equation}
In Figure \ref{fig:enter-label}, points $P_1^\pm$ and $P_2^\pm$ are shown where conditions (\ref{Condition 0+}), (\ref{Conditions 1- z}) and   (\ref{Conditions 1- y}) are simultaneously satisfied with $\nu = 1$. For the points $P_1^\pm$, $m = -2i$ was used, meaning that $M$ is within the interval given by (\ref{Condition M}), and for the points $P_2^\pm$, $m = 0.5i$ was used, which corresponds to the degenerate case.

\subsection{Evolution of the initial state}
Since $|\psi_0^{+}\rangle$ and $|\psi_1^{-}\rangle$ are eigenstates of the Hamiltonian \eqref{Hamiltonian of the trapped ion 3}, we will have that the evolution of the sum of these two eigenfunctions will be
\begin{equation}
    |\psi(t)\rangle=e^{-it\hat{H}_{ion}}|\psi(0)\rangle=\frac{1}{N}\left[c_1 e^{-it E_0^{+}}|\psi_0^{+}\rangle + c_2 e^{-it E_1^{-}}|\psi_1^{-}\rangle \right],
    \label{state ket 0+ 1-}
\end{equation}
where $|c_1|^2+|c_2|^2=1$ and the relations in \eqref{Solutions 0+ 1- without considering the physical properties of the parameters 1} represent the conditions under which the initial state may evolve. Since we know the state ket that describes the system at any given time, we are therefore able to determine the complete dynamics of the trapped ion.
\subsection{Atomic inversion}
\begin{figure}[h!]
    \centering
    \includegraphics[width=.7\linewidth]{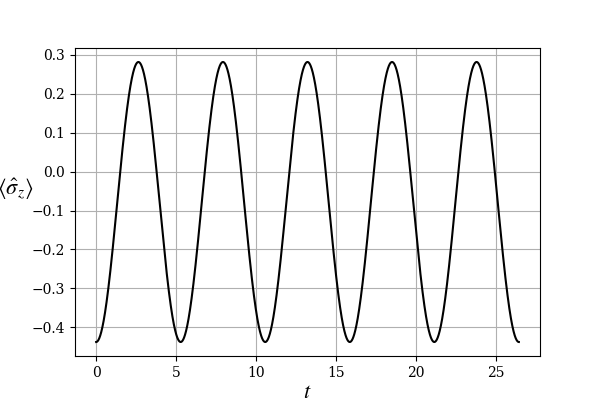}
    \caption{Atomic inversion corresponding to the state $|\psi(t)\rangle$ from the equation (\ref{state ket 0+ 1-}). The parameter values were $(\nu,\Omega,\eta,\delta,m,d_0,d_1,c_1,c_2)=(1.0,-0.5,-0.75,-0.1875,-2i,1,-2i,0.7071,0.7071)$.}
    \label{fig:Atomic inversion m=-2i}
\end{figure}

Making use of the state \eqref{state ket 0+ 1-}, we found that the atomic inversion is given by
\begin{multline}
    \langle\psi(t)|\hat{\sigma}_z|\psi(t)\rangle= \frac{1}{\left| N \right|^2}     \left[ \left| \frac{c_1}{N_0^+} \right|^2 \left( -1 + \left( \frac{\Omega}{\nu} \right)^2 + \left( \frac{\nu \eta}{\Omega} \right)^2 \right)  \right. \\ \left. +\left| \frac{c_2}{N_1^-} \right|^2 \left( \left| d_0 \right|^2 \left( 1 - \left( \frac{\Omega}{2 \nu} \right)^2 \right) + \left| d_1 \right|^2 \left( 1 - \left( \frac{\Omega}{\nu} \right)^2 - 2 \left( \frac{\nu \eta}{\Omega} \right)^2 \right) \right) \right. \\ \left.+ 2e^{-\eta^2/2}   \text{Re} \left( \frac{c_1^* c_2   }{N_{0}^{+*} N_{1}^-}e^{i t (E_{0}^+ - E_{1}^-)}    \left( d_0 \left( \frac{\Omega}{2 \nu} + \frac{\nu \eta^2}{\Omega} \right) - i \eta d_1\frac{\nu}{\Omega} \right) \right) \right].
    \label{Atomic inversion 0+ 1-}
\end{multline}
Figure \ref{fig:Atomic inversion m=-2i}  shows the atomic inversion corresponding to the state $|\psi(t)\rangle$ from the equation (\ref{state ket 0+ 1-}). The case in which the vibrational frequency is twice the magnitude of the field intensity was considered, that is, the system was studied in the medium intensity regime with $\nu = 1$ and $\Omega = -0.5$. Rabi oscillations are observed; however, the frequency is not given by the Rabi frequency $\Omega$. Instead, in general, in the low and medium intensity regimes, it is found that the frequency is given by the difference between the eigenenergies of the eigenstates, $E_0^+ - E_1^-=\delta-\nu$. In the medium intensity regime, if $|\Omega| \approx \nu$, then according to the second equation in (\ref{Solutions 0+ 1- without considering the physical properties of the parameters 1}), $\eta \ll 1$, which indicates that we are in the Lamb-Dicke regime. From the third equation in (\ref{Solutions 0+ 1- without considering the physical properties of the parameters 1}), it is deduced that $\delta \rightarrow 0$, and therefore the oscillation frequency of the atomic inversion will be given by the Rabi frequency $\Omega$. In the low intensity regime, $|\Omega| \ll \nu$, it is found from Eq. (\ref{Solutions 0+ 1- without considering the physical properties of the parameters 1}) that $\eta = \pm \frac{\sqrt{3}}{2}$, $\delta = -\frac{\nu}{4}$, and $m = \pm \sqrt{3}i$. Therefore, the Rabi oscillations will have a frequency given by $-\frac{5}{4}\nu$.
 
\newpage
\subsection{Average number of phonons}
\begin{figure}[h!]
    \centering
    \includegraphics[width=.7\linewidth]{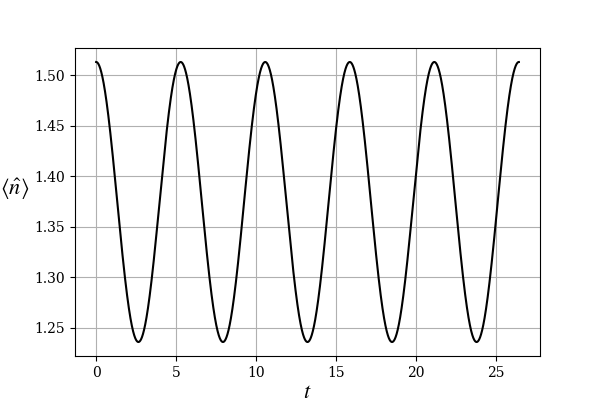}
    \caption{Average number of phonons corresponding to the state $|\psi(t)\rangle$ from the equation \eqref{state ket 0+ 1-}. The parameter values are $(\nu,\Omega,\eta,\delta,m,d_0,d_1,c_1,c_2)=(1.0,-0.5,-0.75,-0.1875,-2i,1,-2i,0.7071,0.7071)$.}
    \label{fig:Average number of phonons m=-2i}
\end{figure}

The average phonon number of the ion's vibrational motion is given by
\begin{multline}
    \langle\psi(t)|\hat{n}|\psi(t)\rangle=   \frac{1}{|N|^2} \left[  \left|\frac{c_1}{N_0^+}\right|^2 \eta^2 \left(1 + \left(\frac{\nu}{\Omega}\right)^2 \right)\right. \\ \left.  +  \left|\frac{c_2}{N_1^-}\right|^2 \left(|d_0|^2 \eta^2 + |d_1|^2 \left(1 + \eta^2 + \left(\frac{\Omega}{\nu}\right)^2 + \left(2\eta \frac{\nu}{\Omega}\right)^2 \right) + i \eta \left( d_0 d_1^* - d_0^* d_1 \right) \right)  \right.  \\
 \left. + 2e^{-\eta^2/2}   \text{Re} \left\lbrace \frac{c_1^* c_2   }{N_{0}^{+*} N_{1}^-}e^{i t (E_{0}^+ - E_{1}^-)}   \left( d_0 \frac{\nu}{\Omega} \eta^2 + i \eta d_1 \left( \frac{\Omega}{\nu} - \frac{\nu}{\Omega} + 3 \frac{\nu}{\Omega} \eta^2 \right) \right) \right\rbrace \right].
 \label{Average number of phonons 0+ 1-}
\end{multline}
Figure \ref{fig:Average number of phonons m=-2i} shows the time evolution of the expected value of the phonon number for the ion's vibrational motion. It has been considered that $\nu = 1$ and $\Omega = -0.5$. As in the case of atomic inversion, a periodic behavior with frequency $E_0^+ - E_1^-$ is observed.

In Figures (\ref{fig:Atomic inversion m=-2i}) and (\ref{fig:Average number of phonons m=-2i}), the medium intensity regime was considered, and it is observed that there is a periodic exchange of energy between the vibrational mode and the internal energy of the ion,  which is typical behavior when the rotating terms of the trapped ion Hamiltonian dominate over the counter-rotating terms.

\section{Conclusions}
Two {\it entangled} eigenstates of the ion-laser Hamiltonian were used to produce dynamical properties. They are chosen from different sets of families. so that they have different eigenvalues (which allows the dynamics) and in such a way that the conditions for their simultaneous existence are fulfilled. In particular, it was shown that the atomic inversion and the average number of phonons undergo Rabi oscillations, with a frequency given by the difference between the eigenenergies of the eigenstates. The analysis was made by doing only the optical rotating wave approximation but keeping the counter rotating vibrational terms. The necessary conditions for the evolution of the proposed initial state indicate that the system's dynamics could occur in the low intensity regime or in the medium intensity regime.

\end{document}